\documentclass[%
prl,
frontmatterverbose, 
showpacs,
 amsmath,amssymb,
 aps,
floatfix,
twocolumn
]{revtex4-1}

\usepackage{graphicx}
\usepackage{dcolumn}
\usepackage{bm}

\usepackage{pstricks}
\usepackage{enumerate}
\usepackage{amssymb,amsmath,latexsym}
\usepackage{amsthm}
\usepackage[plainpages=false]{hyperref}
\usepackage{palatino}
\usepackage{pstricks-add}
\usepackage{pst-plot}
\usepackage{pst-xkey}
\usepackage{multirow}
\usepackage{slashbox}
\usepackage{dsfont}
\usepackage{pifont}
\usepackage{bm}
\usepackage{textcomp}
\usepackage{microtype}
\usepackage{url}
\usepackage{units}
\usepackage{hyperref}
\usepackage{breakurl}

\newcommand{\st}{\operatorname{s.t.}}

\begin{document}

\title{Nonlocality Is Transitive}

\author{Sandro Coretti}
\affiliation{Computer Science Department, ETH Zurich, CH-8092 Z\"urich, Switzerland}

\author{Esther H{\"a}nggi}
\affiliation{Centre for Quantum Technologies, National University of Singapore, Singapore}

\author{Stefan Wolf}
\affiliation{Computer Science Department, ETH Zurich, CH-8092 Z\"urich, Switzerland }

\date{September 21, 2011}

\begin{abstract}
We show 
a transitivity property of nonlocal correlations: 
There exist tripartite nonsignaling correlations of which the bipartite 
marginals between $A$ and $B$ as well as $B$ and $C$ are nonlocal and any 
  tripartite nonsignaling system between $A$, $B$, and $C$ consistent with them must be 
such that the bipartite marginal between $A$ and $C$ is also nonlocal.  
This property represents a step towards ruling out certain 
alternative models 
for the explanation of quantum correlations such as  hidden communication at finite speed.
Whereas it is not possible to rule out this model experimentally, it is the goal of our 
approach to demonstrate this explanation to 
be logically inconsistent: either the communication cannot remain hidden, or its speed has to 
be infinite. 
The existence of a three-party system that is pairwise nonlocal is of independent interest in the 
light of the monogamy property of nonlocality.

\end{abstract}

\maketitle


\paragraph*{Introduction.}|\,
In a classical world, correlations between distant observers are either due 
to preshared information or communication~\cite{gisin}. Shared quantum 
information, i.e., measurements on two (or more) distant parts of an 
entangled quantum state, however, can lead to correlations which are, on one 
hand, stronger 
than what can be achieved by shared information, but, on the other hand, do 
not allow for  
communication. These correlations are called 
nonlocal~\cite{epr,bellinequality}, and the formalism of quantum physics 
predicts that they occur no matter the distance between the observers nor the 
position in space-time they perform their measurements at. 

While quantum theory is  well-established  and  no 
experiment has contradicted its predictions, the question remains open whether there  
could be other physical theories describing our world. The fact that 
correlations predicted by quantum theory are nonlocal and occur even when 
measurements are performed in a spacelike separated way~\cite{aspect} 
implies that such a theory cannot be limited to local hidden 
variables~\cite{bellinequality} and  communication at the 
speed of light. Among alternative  models which could explain the 
experimental observations, one proposition is to consider a physical theory 
based on local hidden variables augmented by  
superluminal hidden communication (in a preferred reference frame) 
for transmitting the nonlocal correlations. If this superluminal 
communication were 
 of infinite speed, this model would be consistent with the 
predictions of quantum theory; however,  is it possible that it occurs at 
some 
finite speed (possibly much faster than the speed of light)? 
It has been pointed out that such communication alone is insufficient~\cite{ScaraniA,ScaraniB}; but what if
it is augmented with hidden variables?  

If the cause of nonlocal correlations was hidden communication 
at finite speed, then the correlations could only be observed between two 
observers as long as the hidden communication can
 travel from one to the other; in case the observers measure their systems 
``too simultaneously,'' their correlations would have to turn local. 
Experiments can, therefore, give a lower bound on the required speed of 
such a communication in a specific reference frame~\cite{experiment};   
however, they can never exclude that the hidden communication occurred at 
an even higher, but still finite, speed. Hence, no 
experiment can rule out any finite speed. 

\begin{figure}[t]
\centering
\includegraphics[width=7.5cm]{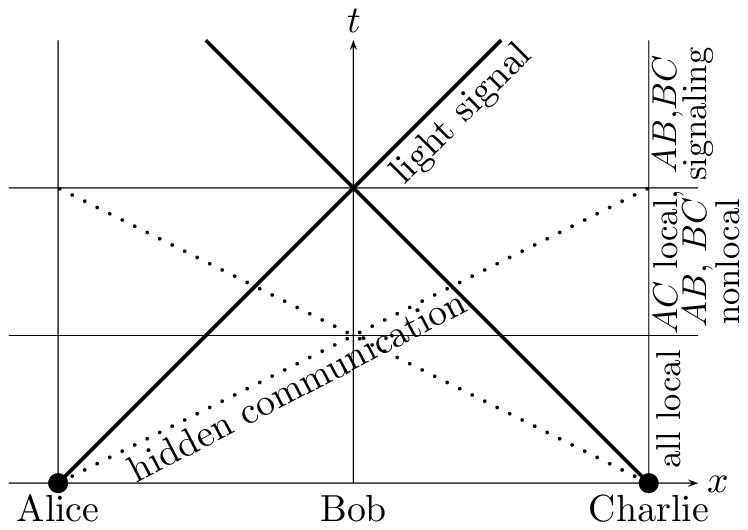}
\caption{
The correlations between Alice, Bob, and Charlie, if Alice and 
Charlie measure their system at $t=0$. Depending on when Bob performs his 
measurement, his correlation with them can be either: local (before the 
hidden communication arrived), nonlocal but nonsignaling (the hidden 
communication but no signal arrived), signaling (light signal arrived). 
Our result states that there are nonlocal correlations between Alice and Bob 
and between Bob and Charlie which, if they could be observed, would exclude 
the existence of the sector where only $AB$ and $BC$ are nonlocal, 
whereas $AC$ is local.
\label{fig:tripartite}}
\end{figure}

For that reason, it is our goal to 
rule out this model in principle, by showing that the assumptions that 
the communication is both hidden and of finite speed lead to a logical 
contradiction. 
This can be done by a Gedankenexperiment~\cite{ScaraniA,ScaraniB}, for which we 
need to find  
correlations between three parties Alice, Bob, and Charlie that are  ``transitive'' 
in the following way: Assume 
Alice and Bob as well as Bob and Charlie are both close enough for the hidden communication to arrive, i.e., 
the correlations $AB$ and $BC$ are nonlocal, while Alice and Charlie are far apart 
(see Figure~\ref{fig:tripartite})~\footnote{Because of the nonsignaling 
condition, $AB$ and $BC$ must then correspond to the marginals of the tripartite system~\cite{ScaraniB}.
}. 
Now, if the marginal bipartite correlations $AB$ and $BC$ are such that any 
consistent nonsignaling correlation must also be nonlocal between Alice and Charlie, 
then the speed of any hidden communication would necessarily have to be infinite.
Therefore, this ``transitivity of nonlocality'' 
rules out finite-speed communication as its explanation in principle, 
independently of any possible experiment.

Note that under the assumption that the correlations are obtained  by 
measurements on a quantum state~$\rho_{ABC}$, nonlocality between Alice and Charlie 
can be inferred even from correlations between Alice and Bob ($\rho_{AB}$) and between 
Bob and Charlie ($\rho_{BC}$)  which are local. The reason is that the bipartite marginals of a 
state (almost always) determine the full state~\cite{LindenWootters}. However, 
since it is our goal to 
compare quantum physics to alternative models, we must reason beyond the 
quantum formalism, i.e., in terms of 
input-output systems 
instead of quantum states. 

In summary, it is our goal to find nonlocal correlations $AB$ and $BC$ that  imply 
nonlocality between $AC$ under any nonsignaling 
composition. 
We show that such correlations, which we call transitive nonlocal, do 
indeed exist, even with respect to Bell inequalities with as little as four measurement 
settings and two outcomes. It is an open question whether the correlations we 
describe are consistent with quantum theory and whether we could find  a tripartite quantum state $\rho_{ABC}$ and 
measurements which lead to the input-output systems of $AB$ and of $BC$ that 
imply nonlocality between $A$ and $C$. This would rule out any 
explanation of 
quantum-physically achievable nonlocality based on 
 finite-speed 
hidden-communication. 

\paragraph*{Preliminaries.}|\,
We characterize a tripartite (physical) system by the probabilities of the 
measurement results given the choice of measurement, i.e., a conditional 
probability distribution $P_{\mathrm{XYZ}|\mathrm{UVW}}$, where $U$ and $X$ are 
Alice's choice of measurement and measurement result, respectively, 
and similarly, $V$ and 
$Y$ are associated with Bob, and $W$ and $Z$ with Charlie. The system 
$P_{\mathrm{XYZ}|\mathrm{UVW}}$ is called nonsignaling if by interacting with any 
marginal side no information about the choice of measurement at the other 
side(s) can be obtained. For example, it is nonsignaling from Charlie to 
Alice and Bob if for all $x$, $y$, $u$, $v$, $w$, $w^\prime$ it holds that
\begin{multline}
\label{eq:ns} \sum\nolimits_z P_{\mathrm{XYZ}|\mathrm{UVW}}(x,y,z,u,v,w)=\\
 \sum\nolimits_z P_{\mathrm{XYZ}|\mathrm{UVW}}(x,y,z,u,v,w^\prime)\;.
 \end{multline}
We require the system to be nonsignaling between all possible two disjoint 
subsets of Alice, Bob, and Charlie. 

A system is called local deterministic if the output on each side is a 
deterministic function on the input only on this side, i.e.,
\begin{align}
\nonumber P_{\mathrm{XYZ}|\mathrm{UVW}}(x,y,z,u,v,w)&
= \delta_{f(u)x}\cdot \delta_{g(v)y}
 \cdot \delta_{h(w)z}\; , 
\end{align}
where $f\colon \mathcal{U}\mapsto \mathcal{X}$ is a function mapping each 
input to a fixed output, and similarly for $g$ and $h$. A system is called 
local if it is a convex combination of local deterministic systems. 
Local systems are exactly the ones which can be described by local hidden 
variables. A system which is nonsignaling, but not local, is called 
nonlocal.

Note that both the space of nonsignaling as well as the space of 
local systems form a convex polytope (for any  number of inputs and 
outputs). The polytope of local systems is (in general) strictly contained in 
the space of nonsignaling systems. 

{B}ell inequalities~\cite{bellinequality,CHSH} are  linear in the 
probabilities $P_{\mathrm{XYZ}|\mathrm{UVW}}(x,y,z,u,v,w)$ and 
fulfilled by any local system. 
If we write $\vec{p}$ for the vector where the entries are all conditional 
probabilities $P_{\mathrm{XYZ}|\mathrm{UVW}}(x,y,z,u,v,w)$, a 
{B}ell inequality is of the form 
$\vec{b}^\intercal\,  \vec{p}\leq c$, where $\vec{b}$ contains the linear 
coefficients describing the Bell inequality, and $c$ is a scalar. 
For example, the  half-spaces determining the local polytope are {B}ell 
inequalities. Conversely, this also implies that any nonlocal 
system must necessarily violate some Bell inequality. For a system 
$P_{\mathrm{XYZ}|\mathrm{UVW}}(x,y,z,u,v,w)$  (not necessarily local) described by 
$\vec{p}^{\prime}$, we will say that it reaches a {B}ell value of 
$c^{\prime}$ if $\vec{b}^\intercal\; \vec{p}^{\prime}= c^{\prime}$. 

For a given system $P_{\mathrm{XYZ}|\mathrm{UVW}}(x,y,z,u,v,w)$, the question whether this 
system is local can be cast as a linear-programming problem, 
i.e., an optimization problem where the objective function is a linear 
function of some vector $\vec{x}$, and the constraints are linear 
equalities or inequalities in $\vec{x}$ (see, e.g.,~\cite{lp} for a good 
introduction to linear programming). More precisely, by solving
\begin{align}
\nonumber \max:\ &\sum\nolimits_i q_i & 
\nonumber \st:\ &A\cdot \vec{q} \leq \vec{p} &
\nonumber &q_i\geq 0\ \text{ for all }i\;, 
\end{align}
where $q_i$ are the entries of the vector $\vec{q}$ to be optimized over, and 
the columns of $A$ are all possible local deterministic systems of this 
number of inputs and outputs. If the optimal value is $1$, then the system is local, 
if it is smaller than $1$, it is nonlocal~\footnote{The linear programs we 
consider here are all necessarily feasible and their optimal value is 
finite. In this case, the optimal solution can be found using 
numerical optimization techniques and once a solution is found, a certificate 
of its optimality can be given.} (see also~\cite{localpart}). 

Additionally, note that the nonsignaling conditions (\ref{eq:ns}) are linear 
in the probabilities, more precisely of the form $A_{\mathrm{n-s}}\, 
\vec{p}=\vec{0}$. The same holds for the conditions  defining a  probability 
distribution, i.e., normalization ($A_{\mathrm{norm}}\, \vec{p}={1}$) and 
positivity (${p}_i\geq {0}$ for all $i$). This implies that the maximum or 
minimum Bell values reachable by a nonsignaling (or local) system can  be 
calculated by a linear program. For example, the maximum {B}ell value 
reachable by a nonsignaling system corresponds to~\cite{toner}
\begin{align}
\nonumber \max:\ &{\vec{b}}^\intercal \cdot \vec{p} \\
\nonumber \st:\ & A_{\mathrm{n-s}}\cdot \vec{p}=\vec{0} &
\nonumber &A_{\mathrm{norm}}\cdot \vec{p}={1} &
\nonumber &p_i\geq 0\ \text{ for all }i\;.
\end{align}
On the other hand, we can minimize the {B}ell value consistent with 
certain constraints (such as, for example, a fixed marginal) and, therefore, 
test whether these constraints are sufficient to imply a {B}ell inequality violation.

\paragraph*{Transitivity.}|\,
In order to find a system that is transitive nonlocal, we will use the 
{B}ell inequalities given in~\cite{brunner,palvertesi} as candidates. Note 
that the best-known Bell inequality --- and the only  one for the case of two 
inputs and outputs --- the {CHSH} inequality~\cite{CHSH} is 
monogamous~\cite{toner}, i.e., a nonsignaling system which violates it 
between Alice and Bob cannot at the same time violate it  between Bob and 
Charlie. Consequently, we need to consider {B}ell inequalities with a larger 
number of inputs and/or outputs. In the following, this will be {B}ell 
inequalities with binary outcomes but with up to four inputs. 

To find a tripartite system which we can then test for transitivity of 
nonlocality, we  proceed as follows. We choose two Bell inequalities, which 
Alice and Bob as well as Bob and Charlie should violate. We then maximize the 
sum of the values of these two Bell inequalities twice, subject to the 
following constraints: 
{(i)}~Alice, Bob, and Charlie share a tripartite nonsignaling system.
{(ii)}~Alice, Bob, and Charlie share a tripartite nonsignaling system of 
which the marginal of Alice and Charlie is local.
If the optimal value obtained in the first optimization is higher than the 
one obtained in the second optimization, the tripartite system giving rise to 
this value cannot be local between Alice and Charlie. A complete list 
of Bell inequalities which have been tested using this approach and  imply 
transitivity of nonlocality can be found in~\cite{sandrothesis}. 

\begin{figure}[htp!]
    \centering
\includegraphics[width=8.6cm]{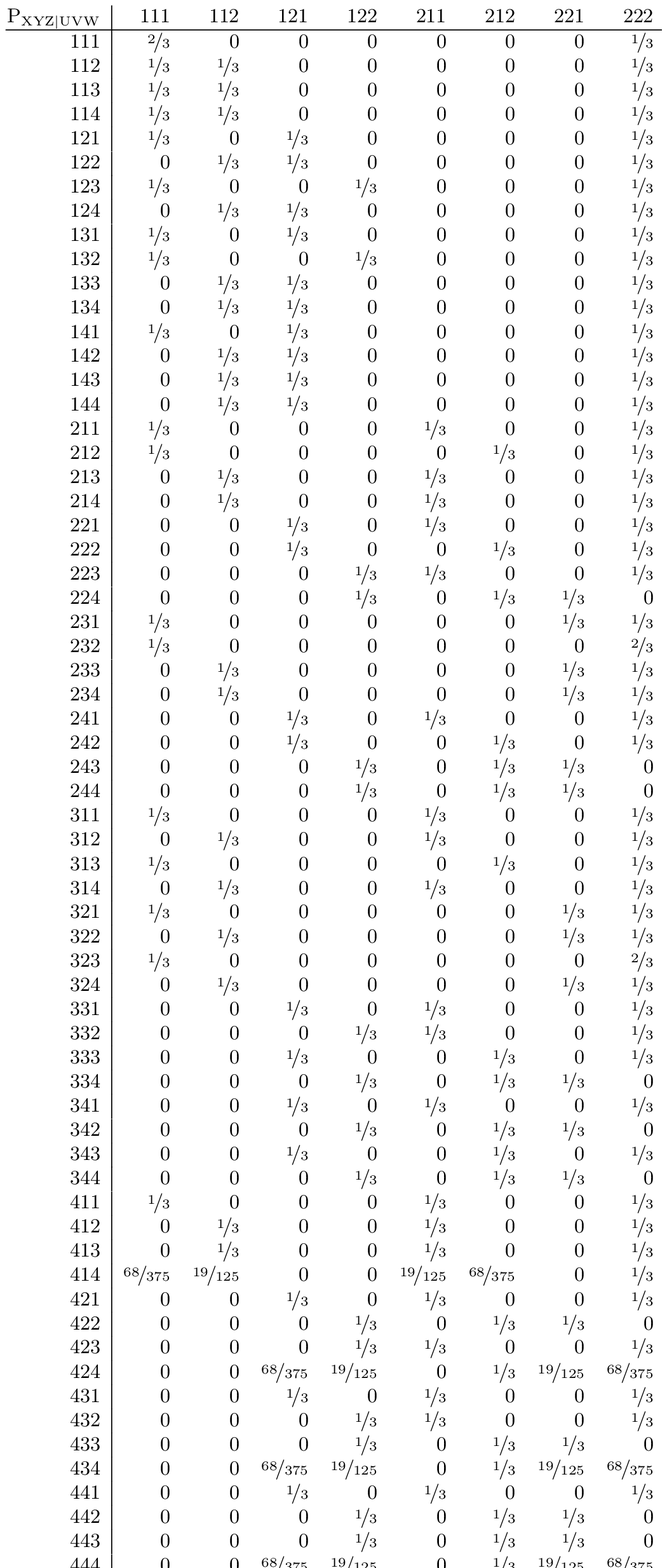}
\caption{\label{table:system} A tripartite nonsignaling system 
$P_{XYZ|UVW}$. The rows contain the different possible inputs of Alice, Bob 
and Charlie and the columns the outputs.}
  \end{figure}

The above approach tells us when a system between Alice and Charlie must be 
nonlocal, i.e., must violate some Bell inequality. It does not 
necessarily imply that there is a specific Bell inequality which must  
be violated between Alice and Charlie. 
Nevertheless, we can check whether this is the case by taking the marginal 
systems of Alice and Bob and Bob and Charlie obtained from the first 
optimization above, and then minimize the Bell value of any tripartite 
nonsignaling system consistent with these marginals. 
An example of such a tripartite 
nonsignaling system which must even violate a specific Bell 
inequality is given in Figure~\ref{table:system} (see also~\cite{sandrothesis}). 

\newpage
Figure~\ref{table:system} describes a tripartite nonsignaling system. 
Consider  the Bell inequalities $I_{4422}^{11}$ and $I_{4422}^{3}$ 
from~\cite{brunner} determined by the following coefficients. 
  \begin{displaymath}
    \begin{array}{ r | r r r r}
      I_{4422}^{11} & -2 & -1 & -1 & 0 \\
      \hline 
      -2 & 1 & 1 & 1 & 2 \\
      -1 & 1 & 0 & 2 & -1 \\
      -1 & 1 & 2 & -1 & -1 \\
      0 & 2 & -1 & -1 & -1 \\
    \end{array}
    \leq 0\;,
\end{displaymath}
\begin{displaymath}
    \begin{array}{ r |rrrr}
      I_{4422}^{3} & -2 & -1 & -1 & 0\\
      \hline 
      -1 & 1 & 1 & 1 & 1 \\
      0 & 0 & 1 & 0 & -1 \\
      0 & 1 & -1 & 1 & -1 \\
      0 & 1 & 0 & -1 & 0 \\
    \end{array}
    \leq 0\,.
  \end{displaymath}
 We used here the same notation as~\cite{brunner} to describe a 
  Bell inequality of a bipartite system with binary outputs and $m$ inputs, i.e., the table gives the coefficients associated with the probabilities of the first output (which we denote here by $1$) of a bipartite system $P_{\mathrm{XY}|\mathrm{UV}}$ in the following way.
  \begin{displaymath} 
        \begin{array}{ l | lll }
        I & P_{\mathrm{Y}|\mathrm{V}}(1,1) &  \cdots & P_{\mathrm{Y}|\mathrm{V}}(1,m) \\
        \hline 
        P_{\mathrm{X}|\mathrm{U}}(1,1) & P_{\mathrm{XY}|\mathrm{UV}}(1,1,1,1) & \cdots  & P_{\mathrm{XY}|\mathrm{UV}}(1,1,1,m)  \\
        \vdots & \vdots & \ddots & \vdots  \\
        P_{\mathrm{X}|\mathrm{U}}(1,m) & P_{\mathrm{XY}|\mathrm{UV}}(1,1,m,1) & \cdots & P_{\mathrm{XY}|\mathrm{UV}}(1,1,m,m) \\
      \end{array}
  \end{displaymath}
It follows by a straight-forward calculation  that the system given in 
Figure~\ref{table:system} violates $I_{4422}^{11}$ for the bipartite 
marginals $AB$ and $BC$ with a value of $2/3$ each. 
Additionally, $AC$ violates $I_{4422}^{3}$ reaching a value of 
$1/3$. Minimizing this value for any nonsignaling system consistent with the 
marginals $AB$ and $BC$, as obtained from Figure~\ref{table:system}, shows 
that this is at the same time the minimal value which can be reached. The 
system given in Figure~\ref{table:system} is, therefore, transitive 
nonlocal with 
respect to the Bell inequality $I_{4422}^{3}$.

\paragraph*{Concluding remarks and open questions.}|\,
Measurements on entangled quantum systems
can lead to correlations
which ask for explanations. Possible such explanations are 
shared information (so-called hidden variables) or some sort of
communication (which would need to be faster than the speed
of light, as experiments have indicated --- so-called hidden
communication). It has been shown by Bell and by Gisin and
Scarani, respectively, that one of these two resources  is
insufficient to explain the correlations in general. We provide
strong evidence that this 
even holds for both combined. More specifically,
we show that nonlocal correlations can have some sort of
transitivity property: There exist pairs of bipartite correlations
between $AB$ and between $BC$ --- with identical marginal
behavior in $B$ --- such that any composition thereof to a 
three-party nonsignaling system $ABC$ must
be such that $A$ and $C$ also share nonlocal correlations. 
This is incompatible with models where nonlocality is transmitted by 
finite-speed hidden communication --- whatever this speed might be ---
as well as models where such correlations 
exist only up to certain distances. The reason is that such models predict situations 
where $AB$ and $BC$ are nonlocal, but $AC$ is local --- for example, if 
$A$ and $C$ measure simultaneously but $B$ measures later. 
We believe that the existence of a three-party
system displaying pairwise nonlocality (and of a Bell inequality allowing 
for this) is of independent interest because of the monogamy property
of nonlocality. 

It is an open question whether the
correlations we consider are quantum-physically realizable, and,
in particular, whether there exists a tripartite quantum state
$\rho_{ABC}$
whose bipartite marginal systems allow for carrying out a similar
reasoning.

We thank Daniel Burgarth, Nicolas Gisin, and Valerio Scarani for helpful 
discussions. 
EH and SW acknowledge support from 
the Swiss National Science Foundation and an ETHIIRA grant of ETH's research 
commission. 
EH acknowledges support from  the National Research Foundation (Singapore), and the
Ministry of Education (Singapore). 
Part of this work was done while EH was with ETH Zurich.

\bibliographystyle{apsrev}
\bibliography{tripartite}

\end{document}